\documentclass{llncs}
\input{psfig.sty}
\usepackage {graphicx}
\usepackage{amsmath}

\begin{document}

\mainmatter

\title{A light-based device for solving \\the Hamiltonian path problem}

\author{Mihai Oltean}
\institute{Department of Computer Science,\\
Faculty of Mathematics and Computer Science,\\
Babe\c s-Bolyai University, Kog\u alniceanu 1\\
Cluj-Napoca, 400084, Romania.\\
\email{moltean@cs.ubbcluj.ro}\\
\email{www.cs.ubbcluj.ro/$\sim$moltean}
}

\maketitle

\begin{abstract}

In this paper we suggest the use of light for performing useful computations. Namely, we propose a special device which uses light rays for solving the Hamiltonian path problem on a directed graph. The device has a graph-like representation and the light is traversing it following the routes given by the connections between nodes. In each node the rays are uniquely marked so that they can be easily identified. At the destination node we will search only for particular rays that have passed only once through each node. We show that the proposed device can solve small and medium instances of the problem in reasonable time.

\end{abstract}

\section{Introduction}

Using the light to perform computations is an exciting idea whose applications can be already seen on the market.

An important step was made by Intel researchers who have developed the first continuous wave all-silicon laser using a physical property called the Raman Effect \cite{Faist,Paniccia,rong1,rong2}. The device could lead to such practical applications as optical amplifiers, lasers, wavelength converters, and new kinds of lossless optical devices.

Another solution comes from Lenslet \cite{lenslet} which has created a very fast processor for vector-matrix multiplications. This processor can perform up to 8000 Giga Multiple-Accumulate instructions per second. Lenslet technology has already been applied to data analysis using k-mean algorithm \cite{kmean} and video compression.

In this paper we suggest a new way of performing computations by using some properties of light waves. The idea may be used within a special device for solving the Hamiltonian path problem.

We are not taking into account the quantum properties of light which have been used for solving the Traveling Salesman Problem \cite{cerny,greenwood}.

The paper is organized as follows: The Hamiltonian path problem is briefly described in section \ref{HP}. The proposed device is presented in section \ref{proposed}. Mathematical background of the labeling system is described in section \ref{math}. The way in which the proposed device works is given in section \ref{howorks}. A list of components required by the proposed device is given in section \ref{hard}. Complexity is discussed in section \ref{complexity}. Suggestions for improving the device are given in sections \ref{speed_reduce} and \ref{particular_graphs}. Further work directions are suggested in section \ref{further}.

\section{The Hamiltonian path problem}
\label{HP}

The description of the Hamiltonian Path (HP) problem for a direct graph is the following:

Given a directed graph $G = (V, E)$ with $|V| = n$ nodes and a start node ($v_{start}$) and a stop node ($v_{stop}$), the problem asks to compute is there is a simple path, beginning with node $v_{start}$ and ending with node $v_{stop}$, containing all nodes exactly once. The output for this decision problem is either YES or NO depending on whether the Hamiltonian path does exist or not.

The Hamiltonian path problem arises in many real-word applications \cite{Ascheuer,Doniach}.

The problem belongs to the class of NP-complete problems \cite{garey}. No polynomial time algorithm is known for it. 

A small instance of this problem was also the first problem solved using a DNA computer \cite{adleman}.

\section{The proposed device}
\label{proposed}

Our idea is based on two properties of light:

\begin{itemize}

\item{The speed of light has a limit. The value of the limit is not very important at this stage of explanation. What is important is the fact that we can delay the ray by forcing it to pass through an optical fiber cable of a certain length.}

\item{The ray can be easily divided into multiple rays of smaller intensity/power.}

\end{itemize}

Initially a light ray is sent to the start node. Generally speaking two operations must be performed when a ray passes through a node :

\begin{itemize}

\item{The light ray is marked uniquely so that we know that it has passed through that node.}

\item{The ray is divided into a number of rays equal to the external degree of that node. Each obtained ray is directed toward one of the nodes connected to the current node.}

\end{itemize}

At the destination node we will search only for particular rays that have passed only once through each node.

This section deeply describes the proposed system. First step is to find a way to mark the signals which passes through nodes such that the interesting signals can be easily identified at the destination node. The mathematical background required for this operation is described in section \ref{math} and the hardware implementation of the labeling system is described in section \ref{hard}.

\subsection{Labeling system}
\label{labeling}

At the destination node we will wait for a particular ray which has passed through all nodes of the graph exactly once. This is why we need to find a way to label that particular ray so that it could be easily identified. 

Actually we are interested in marking all rays which pass through a particular node with a unique label, such that Hamiltonian path is uniquely identified at the destination node ($v_{stop}$).

In the solution proposed in this paper, the rays passing through a node are marked by delaying them with a certain amount of time. This delay can be easily obtained by forcing the rays to pass through an optical fiber of a certain length. Roughly speaking, we will know if a certain ray has traversed a Hamiltonian path only if its delay (at the destination node) is equal to the sum of delays of all nodes in that graph. We will also know the particular moment when the expected ray (the one which has completed a Hamiltonian path) will arrive. In this case the only thing that we have to do is to "listen" if there is a fluctuation in the intensity of the signal at that particular moment. Due to the special properties of the proposed system we know that no other ray will arrive, at the destination node, at the moment when the Hamiltonian path ray has arrived.

The delays, which are introduced by each node, cannot take any values. If we would put random values for delays we might have different rays (which are not Hamiltonian paths) arriving, at the destination node, in the same time with a ray representing a Hamiltonian path.

We need only the ray, which has traversed a Hamiltonian path, to arrive in the destination node at the moment equal to the sum of delays of each node (the moment when the ray has entered in the start node is considered moment 0). Thus, the delaying system must have the following property:\\

\textbf{Property of the delaying system}

Let us denote by $d_1$, $d_2$, ..., $d_n$ the delays introduced by each node of the graph. A correct set of values for this system must satisfy the condition:

$d_1+d_2+...+d_n \not= a_1 \cdot d_1+a_2 \cdot d_2+...+a_n \cdot d_n$,

where $a_i$ ($1 \leq i \leq n$) are natural numbers and cannot be all 1 in the same time. 

If a given value $a_i$ is strictly greater than 1 it means that the ray has passed at least twice through node 1.

\subsection{Mathematical background for the labeling system}
\label{math}

Finding the appropriate labeling system was two steps process. First of all we have written a computer program which generates this numbers by using a backtracking procedure \cite{cormen1}. We also wanted to generate numbers such that the highest number in a system is the smallest possible. This will ensure that the network is constructed in an efficient way. The labeling systems generated by our computer programs are given in Table \ref{tab_labeling_system}.

\begin{table}[htbp]
\begin{center}
\caption{The labeling system generated by our backtracking procedure. First column contains the number of nodes of the graph. The second column represents the labels applied to nodes.}
\label{tab_labeling_system}
\begin{tabular}
{p{50pt}p{100pt}}
\hline
n& 
Labels (delays) \\
\hline
1& 
1 \\
2& 
2, 3 \\
3& 
4, 6, 7 \\
4& 
8, 12, 14, 15 \\
5& 
16, 24, 28, 30, 31 \\
6& 
32, 48, 56, 60, 62, 63 \\
\hline
\end{tabular}
\end{center}
\end{table}

From Table \ref{tab_labeling_system} it can easily seen these numbers follow a general rule:

For a graph with $n$ nodes one of the possible labeling systems is:\\

$2^n-2^{n-1}$, 

$2^n-2^{n-2}$, 

$2^n-2^{n-3}$, 

... ,

$2^n-2^0$. \\

As the second step we have to prove that the property of delaying system (see section \ref{labeling}) holds for this sequence of numbers.

Actually we have to prove that the equality:\\

\begin{equation}
\begin{split}
2^n-2^{n-1} + 2^n-2^{n-2} + 2^n-2^{n-3} + ... + 2^n-2^0 =\\
a_1 \cdot (2^n-2^{n-1}) + a_2 \cdot (2^n-2^{n-2}) + a_3 \cdot (2^n-2^{n-3}) + ... + a_n \cdot (2^n-2^0)
\end{split}
\label{eq1}
\end{equation}

\noindent
\\is not possible unless all $a_i$ are equal to 1.\\

The left part of the equality is:\\

\noindent
$2^n-2^{n-1} + 2^n-2^{n-2} + 2^n-2^{n-3} + ... + 2^n-2^0 =$\\
$n \cdot 2^n - (2^{n-1} + 2^{n-2} + 2^{n-3} + ... + 2^0) =$\\
$n \cdot 2^n - 2^n + 1 =$\\
$(n - 1) \cdot 2^n + 1.$\\

First of all we have to see that the equality does not hold if all $a_i$ numbers are at least 1 and at least one number is strictly greater than 1. If this happens the $2^n$ term will be represented at least $n$ times. But, it must be represented only $n-1$ times (see above).

Thus, if at least one number $a_i$ is strictly greater than 1, it means that other numbers $a_j$ must be 0. We will prove that the equality (1) does not hold in this case too.

As discussed above, if one of the coefficients is 0, at least one of the other coefficients must be strictly greater than 1. For instance, if $a_1$ is set to 0, it means that $a_2$ must be set to 3 in order to compensate the missing $2^{n-1}$ term. This is a direct consequence of the fact that $2^{n-1} = 2 \cdot 2^{n-2}$. Of course, we also have to take into account that $2^{n-2}$ must be represented once. This is why we have to set $a_2$ to value 3.

If $a_2$ is also 0 we have to set $a_3$ to value 7 (we need $4 \cdot 2^{n-3}$ in order to compensate the missing term $2^{n-1}$, we also need $2 \cdot 2^{n-3}$ to compensate the missing term $2^{n-2}$ and, of course, the term $2^{n-3}$ must be represented 1 time).

As a general idea: if a particular term ($2^{n-j}$) is missing (the corresponding coefficient $a_j$ is set to 0), it can  be compensate by setting one of the next coefficients to a value of at least 3. But, a coefficient set to 0 means that the term $2^n$ is missing once, and by setting another coefficient to at least 3 we will get at least 2 extra representations for $2^n$. This will mean that right part of the equation (1) will be at least $(n+1) \cdot 2^n + 1$. But, the left part of the equation is only $(n-1) \cdot 2^n + 1$.

With this we have shown that all $a_i (1 \leq i \leq n)$ must be 1 in order to have equality in equation (1). For any other values of $a_i$ the equality (1) does not hold.

An important question is whether this system is the minimal possible (the biggest number is the minimal possible). A partial answer to this question is given in section \ref{complexity}.

\begin{figure}[htbp]
\centerline{\includegraphics[width=1.79in,height=2.81in]{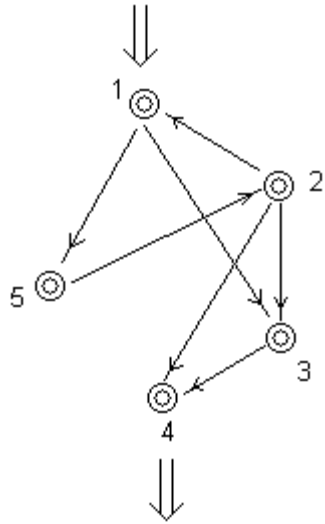}}
\caption{A directed graph with 5 nodes. Start node is 1 and the destination node is 4. The list of arcs is: (1,3), (1, 5), (2, 1), (2, 3), (2, 4), (3, 4), (5, 2)}
\label{graph_fig}
\end{figure}

\subsection{How the system works}
\label{howorks}

An schematic example of a graph-like system is given in Figure \ref{graph_fig}.

In the graph depicted in Figure \ref{graph_fig} the light will enter in node 1. It will be delayed with a certain amount of time and then it will be divided into 2 rays which will be sent to the nodes 3 and 5. In node 3 the ray will be delayed (with the amount of time corresponding to node 3) and then it will be sent to node 4. However, this is not a Hamiltonian path because it has visited only the nodes 1, 3 and 4. The other ray which was sent to node 5 (from node 1) can generate a Hamiltonian path by following the route 1, 5, 2, 3 4.

Note that there is also a cycle in the graph: 1, 5, 2. The cycle will make that some particular rays to be trapped within the system. The rays which have passed once through the previously described cycle are not considered Hamiltonian paths because the moments when they arrive at the destination node are greater than the sum of delays introduced by each node.

The partial paths traversed by rays are given below:\\

1

\hspace{0.25cm} 1, 3

\hspace{0.5cm} 1, 3, 4

\hspace{0.25cm} 1, 5

\hspace{0.5cm} 1, 5, 2

\hspace{0.75cm} 1, 5, 2, 1

\hspace{0.75cm}.............\{paths starting with 1, 5, 2, 1\}

\hspace{0.75cm} 1, 5, 2, 3

\hspace{1cm} 1, 5, 2, 3, 4

\hspace{0.75cm} 1, 5, 2, 4

\subsection{Hardware implementation of the labeling system}
\label{hard}

For implementing the proposed device we need the following components:

\begin{itemize}

\item{a source of light (laser),}

\item{Several beam-splitters for dividing light rays into multiple subrays. A standard beam-splitter is designed using a half-silvered mirror. For dividing a ray into $k$ subrays we need $k-1$ beam-splitters.}

\item{A high speed photodiode for converting light rays into electrical power. The photodiode is placed in the destination node.}

\item{A tool for detecting fluctuations in the intensity of electric power generated by the photodiode (oscilloscope),}

\item{A set of optical fiber cables having certain lengths. These cables are used for connecting nodes and for delaying the signals within nodes. The length of the cables must obey the rules described in section \ref{labeling}. A practical example is given in section \ref{psize}.}

\end{itemize}

\subsection{Complexity}
\label{complexity}

This section answers a very important question: \textit{Why the proposed approach is not a polynomial-time solution for the HP problem?}

At the first sight one may be tempted to say that the proposed approach provides a solution in polynomial time to any instance of the HP problem. The reason behind such claim is given by the ability of the proposed device to provide output to any instance by traversing only once all nodes (O($n$) complexity). This could mean that we have found a polynomial-time algorithm for the HP problem. A direct consequence is obtaining solutions, in polynomial time, for all other NP-Complete problems - since there is a polynomial reduction between them \cite{garey}.

However, this is not our case. As can be seen from Table \ref{tab_labeling_system} the delay time increases exponentially with the number of nodes. Even if the ray has to traverse only $n$ nodes (resulting a complexity of O($n$)), the total time required by the ray to reach the destination node increases exponentially with the number of nodes.

There are two direct consequences which are derived from here:

\begin{itemize}

\item{The length of the optical fibers, used for delaying the signals, increases exponentially with the number of nodes,}

\item{The intensity of the signal decreases exponentially with the number of nodes that are traversed.}

\end{itemize}

These two issues are discussed in sections \ref{psize} and \ref{amplify}.

\subsection{Problem size}
\label{psize}

We are interested in computing the size of the cables required to solve a certain instance of the problem in a small amount of time. This will give as a rough indication on the size of the graphs that can be solved using our system in reasonable time.

This size heavily depends on the accuracy of the measurement tools. The rise-time of the best oscilloscope available on the market is in the range of picoseconds ($10^{-12}$ seconds). This means that we should not have two signals that arrive at 2 consecutive moments at a difference smaller than $10^{-12}$ seconds.

Knowing that the speed of light is $3 \cdot 10^{8} m/s$ we can easily compute the minimal cable length that should be traversed by the ray in order to be delayed with $10^{-12}$ seconds. This is obviously 0.0003 meters.

This value is the minimal delay that should be introduced by a node in order to ensure that the difference between the moments when two consecutive signals arrive at the destination node is greater or equal to the measurable unit of $10^{-12}$ seconds. This will also ensure that we will be able to correctly identify whether the signal has arrived in the destination node at a moment equal to the sum of delays introduced by each node. No other signals will arrive within a range of $10^{-12}$ seconds around that particular moment.

Once we have the length for that minimal delay is quite easy to compute the length of the other cables that are used in order to induce a certain delay.

Recall from section \ref{math}, Table \ref{tab_labeling_system} that a graph with 5 nodes has the following delaying system:\\

16, 24, 28, 30, 31.\\

From the previous reasoning line we have deduced that the smaller indivisible unit is 0.0003. So, we have to multiply these numbers by 0.0003. We obtain: \\

0.00048, 0.00072, 0.00084, 0.0009, 0.00093.\\

These numbers represent the length of the cables that must be used in graph's nodes in order to induce a certain delay.

Note that the delay introduced by the cables connecting the nodes was not taken into account in this example. This is not a limitation of our system. The cables connecting nodes can be set to have some length which must obey the property of delaying system (see section \ref{labeling}). Note that all cables must have the same length. In this case if we have a graph with 4 nodes the length of every cable connecting the nodes must be set to 16 units (the shortest possible - in order to reduce the costs). The length of cables within the nodes should be 24, 28, 30 and 31 units.

The largest length in this sequence is 0.00093 meters. This length is not very big, but for larger graphs the length of the cables within nodes can be a problem.

Once we have the length for that minimal delay is quite easy to compute the maximal number of nodes that a graph can have in order to find the Hamiltonian path in one second. We know the facts:

\begin{itemize}

\item{the largest delay has the form $2^n - 1$ (see equation \ref{eq1}),}
\item{the distance traversed by light in 1 second is $3 \cdot 10^8$ meters,}

\item{the shortest delay possible is 0.0003 meters.}

\end{itemize}

We simply have to solve the equation:

\begin{equation}
2^n \cdot 0.0003 = 3 \cdot 10^8
\end{equation}

This number is about 40 nodes. However, the length of the optic fibers used for inducing the largest delay for this graph is huge: about $8 \cdot 10^{8}$ meters. We cannot expect to have such long cables for our experiments.

However, shorter cables (of several hundreds of kilometers) are already available in the internet networks. They can be easily used for our purpose. Assuming that the longest cable that we have is about 300 kilometers we may solve instances with about 26 nodes. The amount of time required to obtain a solution is about $10^{-6}$ seconds. 

Note that the maximal number of nodes can be increased when the precision of our measurement instruments (oscilloscope and photodiode) is increased. 

Also note that this difficulty is not specific to our system only. Other major unconventional computation paradigms, trying to solve NP-complete problems share the same fate. For instance, a quantity of DNA equal to the mass of Earth is required to solve HP instances of 200 cities using DNA computers \cite{Hartmanis}.

\subsection{Amplifying the signal}
\label{amplify}

Beam splitters are used in our approach for dividing a ray in two or more subrays. Because of that, the intensity of the signal is decreasing. In the worst case we have an exponential decrease of the intensity. For instance, in a graph with $n$ nodes, each signal is divided (within each node) into $n - 1$ signals. Roughly speaking, the intensity of the signal will decrease $n^n$ times.

This means that, at the destination node, we have to be able to detect very small fluctuations in the intensity of the signal. For this purpose we will use a photomultiplier \cite{Flyckt} which is an extremely sensitive detector of light in the ultraviolet, visible and near infrared range. This detector multiplies the signal produced by incident light by as much as $10^8$, from which even single photons can be detected.

\subsection{Improving the device by reducing the speed of the signal}
\label{speed_reduce}

The speed of the light in optic fibers is an important parameter in our device. The problem is that the light is too fast for our measurement tools. We have either to increase the precision of our measurement tools or to decrease the speed of light.

It is known that the speed of light traversing a cable is significantly smaller than the speed of light in the void space. Commercially available cables have limit the speed of the ray wave up to $60\%$ from the original speed of light. This means that we can obtain the same delay by using a shorter cable.

However, this method for reducing the speed of light is not enough. The order of magnitude is still the same. This is why we have the search for other methods for reducing that speed. A very interesting solution was proposed in \cite{Hau,Liu} which is able to reduce the speed of light by 7 orders of magnitude. This could help our mechanism significantly. However, is still a question how to use this idea for our device because of the complex equipment involved in those experiments \cite{Hau,Liu}.

By reducing the speed of light by 7 orders of magnitude we can reduce the size of the involved cables by a similar order. This will help us to solve larger instances of the problem.

\subsection{Improving the performance of the device for particular graphs}
\label{particular_graphs}

The labeling system proposed in section \ref{labeling} is a general one. It can be used for any kind of graph (with any number of nodes and arcs). We have shown that this system has a big problem: the value of the involved numbers increase exponentially with the number of nodes in the graph being solved.

\begin{figure}[htbp]
\centerline{\includegraphics[width=3.2in,height=0.8in]{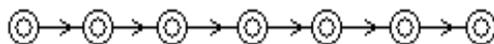}}
\caption{A linear graph with 7 nodes. No delays are required for the nodes of this graph in order to find an Hamiltonian path}
\label{fig_linear_graph}
\end{figure}

But, for particular graphs we can find other labeling systems which are not exponential. For example, the linear graph (see Figure \ref{fig_linear_graph} can be solved by our device by using virtually no delays. In this case the moment when the signal arrives in the destination node is equal to sum of delays introduced by the cables connecting the nodes.

Finding the optimal labeling system for a particular graph is an interesting problem which will be investigated in the near future.

\subsection{Technical challenges}

There are many technical challenges that must be solved when implementing the proposed device. Some of them are:

\begin{itemize}

\item{Cutting the optic fibers to an exact length with high precision. Failing to accomplish this task can lead to errors in detecting a ray which has passed through each node once.}

\item{Finding a high precision oscilloscope. This is an essential step for solving larger instances of the problem (see section \ref{psize}),}

\item{Finding cables long enough so that larger instances of the problem could be solve. This problem might have a simple solution: the internet networks connecting the world cities. It is easy to find cables of hundreds of kilometers connecting distant cities. This will help us to solve instance of more than 10 nodes. However, this solution introduces a difficulty too: cables of certain lengths must be found or the system must be rescaled in order to fit the existing lengths.}

\end{itemize}

\section{Conclusions and further work}
\label{further}

The way in which light can be used for performing useful computations has been suggested in this paper. The techniques are based on the massive parallelism of the light ray.

It has been shown the way in which a light-based device can be used for solving the Hamiltonian path problem. Using the today technology we can build a light-based device which can solve small and medium size instances in several seconds.

Further work directions will be focused on:

\begin{itemize}

\item{Implementing the proposed hardware,}

\item{Finding optimal labeling systems for particular graphs. This will reduce the length of the involved cables significantly,}

\item{Finding other non-trivial problems which can be solved by using the proposed device,}

\item{Finding other ways to introduce delays in the system. The current solution requires cables that are too long and too expensive,}

\item{Using other type of signals instead of light. A possible candidate would be electric power,}

\item{Finding other ways to implement the system of marking the signals which pass through a particular node. This will be very useful because the currently suggested system, based on delays, is too time consuming.}

\end{itemize}

\end{document}